\documentclass[reprint,amsmath,amssymb,superscriptaddress,aps,prl]{revtex4-1}
\usepackage{subfigure}
\usepackage{graphicx}
\usepackage{bm}
\usepackage{color}
\usepackage{times}
\usepackage{ulem}
\newcommand{\super}{($\sqrt{3}\times\sqrt{3}$)\textit{R}30$^{\circ}$ }

\begin{document}
\title{Intrinsic superstructure near atomically clean armchair-type step edge of graphite}
\author{Wenhan Zhang}
\affiliation{Department of Physics and Astronomy, Rutgers University, Piscataway, New Jersey 08854, USA}
\author{Zheng Ju}
\affiliation{Department of Physics and Astronomy, Rutgers University, Piscataway, New Jersey 08854, USA}
\affiliation{Department of physics, Arizona State University, Tempe, Arizona 85287, USA}
\author{Weida Wu}
\email{Corresponding Author.\newline wdwu@physics.rutgers.edu}
\affiliation{Department of Physics and Astronomy, Rutgers University, Piscataway, New Jersey 08854, USA}
\date{\today}

\begin{abstract}
We investigated the electronic superstructure of graphite surface in the vicinity to monoatomic armchair step edges with scanning tunneling microscopy and spectroscopy. Only the \super superstructure is visualized near atomically clean armchair edges, while the honeycomb superstructure is absent. The spectroscopic mapping near the clean armchair edge clearly reveals the \super superstructure on both sides of the step edge. We have also visualized a mixture of \super and honeycomb superstructures near structurally defective armchair edges. Our results suggest that the honeycomb superstructure pattern results from superposition of two sets of \super superstructure with different phases. Our observation solves the mystery of the coexistence of two types of superstructures reported by prior studies.
\end{abstract}

\maketitle
Graphene has drawn extensive attention in material science since its first successful isolation in 2004~\cite{Novoselov2004}. The unconventional electronic properties make it a potental candidate for post-silicon electronic devices~\cite{CastroNeto2009}. Known as a 2D Dirac material, the low-energy electron excitation of graphene behaves like a massless Dirac fermion~\cite{Wehling2014}, which gives rise to novel electronic phenomena, such as the quantum spin Hall effect~\cite{Kane2005PRLGraphene}, Klein tunneling~\cite{Beenakker2008}, and the anomalous quantum Hall effect~\cite{Novoselov2005, Zhang2005}. Recently, superconductivity and magnetism are discovered in twisted bilayer graphene~\cite{Cao2018, Yankowitz2018, Tarnopolsky2019, Sharpe2019}, which provides a new route to the development of superconducting and magnetic devices with carbon-based materials.

\begin{figure}[b]
\centering
\includegraphics[width=0.9\columnwidth]{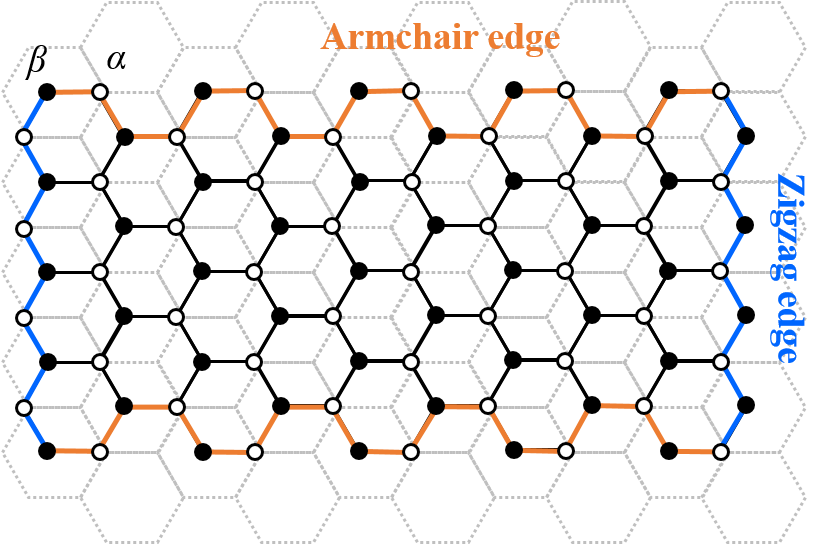}
\caption{(Color online) Schematic of lattice structure of two atomic layers of graphite showing two types of monoatomic step edges. Black (gray) honeycomb lattice represents the upper (lower) atomic layer. The open and closed circles represent $\alpha$- and $\beta$-site carbon atoms of the topmost layer. The zigzag (armchair) edges are denoted by the blue (orange) lines.}
\label{Gr_LatStruc}
\end{figure}

As the graphene is cut into nano-size fractions or semi-infinite sheets, the impact of edge on the electronic properties of graphene becomes significant~\cite{Stein1987, Fujita1996, Nakada1996, Affoune2001, Cancado2004, Enoki2012}. Depending on the edge direction with respect to the lattice vector, there are two simplest types of edges, i.e., armchair and zigzag edges, as illustrated in Figure~\ref{Gr_LatStruc}. They host distinct electronic properties. Based on prior theoretical~\cite{Fujita1996, Nakada1996, Brey2006, Akhmerov2008} and experimental~\cite{Kobayashi2006, Niimi2006, Tao2011, Ziatdinov2013a} studies, there exists an electronic states localized on the zigzag edge of graphene due to the nonbonding $\pi$ electrons of the edge carbon atoms, whereas such edge state is absent on the armchair edge. Such enhanced local density of states (LDOS) on the zigzag edges may give rise to unusual phenomena such as edge magnetism, which are potential for spintronics applications~\cite{Kusakabe2003, Lee2005, Son2006, Yazyev2008}.

In addition, the edge also plays a role of potential barrier and induces electron wave scattering. It is manifested as a superstructure pattern, which can be directly probed by scanning tunneling microscopy. Such superstructure pattern near the step edge on graphite has been extensively studied in the last three decades. However, there is still no consensus on the intrinsic superstructure of ideal step edges. In the vicinity of the armchair edge, some studies show only the \super superstructure~\cite{Giunta2001, Kobayashi2005a, Kobayashi2006}, while others report a coexistence of the \super and honeycomb superstructures~\cite{Niimi2006, Sakai2010}. On the other hand, theoretical calculations predict that there is no superstructure near the zigzag edge~\cite{Giunta2001,Niimi2006}, while some experimental works show signatures of superstructure patterns~\cite{Kobayashi2005a, Ziatdinov2013a}. These discrepancies probably originate from the structural or chemical imperfections of the edge, e.g., mixture of armchair and zigzag edges or extrinsic adsorbates, which hinder the clear observation of intrinsic electronic properties associated with the edges~\cite{Mizes1989, Xhie1991, Shedd1992, Valenzuela-B1995, Ruffieux2000, Lopez2008, Sakai2010, Ziatdinov2014}.

To address this issue, we performed scanning tunneling microscopy and spectroscopy (STM/S) experiments to study clean step edges on the surface of single crystal graphite. This work focuses on the armchair edge, because it is energetically more stable than the zigzag edge~\cite{Lee1997, Kawai2000, Okada2008}. We managed to find atomically clean armchair edges on graphite as long as $\sim30$~nm. In the vicinity of such edge, the \super superstructure is observed, while the honeycomb superstructure is absent. The STS maps measured near the armchair edge show this type of superstructure pattern on both the upper and lower terraces. Interestingly, we visualized a mixture of \super and honeycomb superstructures near structurally defective armchair edge. We propose that the honeycomb superstructure pattern emerges as a superposition of two sets of \super superstructure at their antiphase boundary. Our observation demonstrates that the \super superstructure is intrinsic on the armchair step edge.

We conducted experiments on the single crystals of highly oriented pyrolytic graphite (HOPG). The samples were cleaved \textit{in-situ} in ultra-high vacuum (UHV: $\sim2\times10^{-11}$~mbar) and then immediately transferred into the STM head for measurements. Before cleavage, they were pre-cooled to a low temperature in the STM head, which avoids outgassing of STM head due to a sudden rise of temperature at the moment of sample insertion and thus significantly reduces contamination on the sample surface. The STM measurements in this work was performed at $T=5$~K unless otherwise specified. Electrochemically etched tungsten tips as scanning probes were treated and characterized on single-crystal Au(111) surface~\cite{Chen1998}. The STS mapping measurements were performed with the standard lock-in technique with a modulation frequency $f=455$~Hz and a modulation amplitude $V_\text{mod}=20$~mV. All the step edges of HOPG shown in this work are monoatomic with a step height $\Delta z=3.3$~\AA .

\begin{figure}[ht]
\centering
\includegraphics[width=\columnwidth]{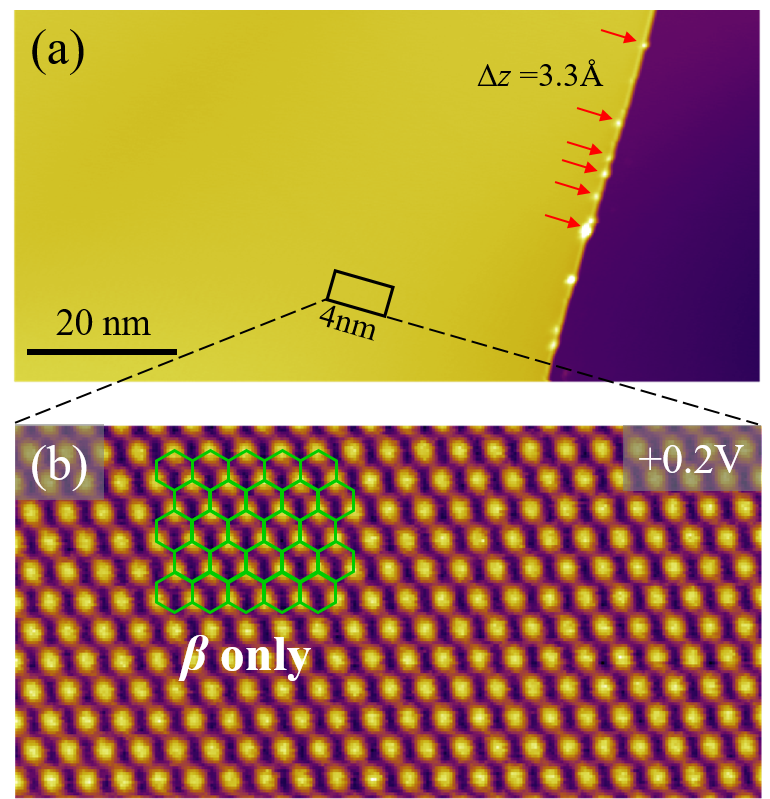}
\caption{(Color online) (a) Large-scale topographic image containing an armchair-type step edge. The red arrows mark representative imperfections such as absorbates on the edge. Tunneling condition: $I=100$~pA, $V=-1$~V. (b) Atomically resolved topographic image taken far from the edge showing intrinsic lattice structure of graphene sheet. Tunneling condition: $I=1$~nA, $V=+0.2$~V. The green honeycomb lattice shows the atomic lattice. Only $\beta$-site carbon atoms are observed.}
\label{Gr_AtomicFlat}
\end{figure}

Fig.~\ref{Gr_AtomicFlat}(a) shows a topographic image near a representative step edge. Such linear step edges can extend over a few hundred nanometers. In the area far from the edge, the intrinsic lattice of topmost layer of graphite is observed, as shown in Fig.~\ref{Gr_AtomicFlat}(b). The atomic corrugation is the most pronounced at low sample bias ($+0.2$~eV) and shows a hexagonal lattice. A phase shift of $2\pi/3$ of such lattice across the step edge indicates that the resolved carbon atoms are at $\beta$ site, i.e., there are no carbon atoms directly below or above in adjacent layers (see Appendix for detailed phase analysis). This is consistent with previous STM studies~\cite{SGwo1993a}. With the atomic resolution, the step edge in Fig.~\ref{Gr_AtomicFlat}(a) can be identified as armchair type, because it is perpendicular to the atomic row of $\beta$-site carbon atoms. The most common monoatomic step edges are armchair type in our measurements. We did not observe linear zigzag edges on naturally cleaved HOPG surfaces. Due to the low-$T$ cleavage, the edge is partially clean. The bright spots marked by red arrows are likely absorbates on the edge, and the straight segments between them are clean and homogeneous.

\begin{figure*}[ht]
\centering
\includegraphics[width=0.9\textwidth]{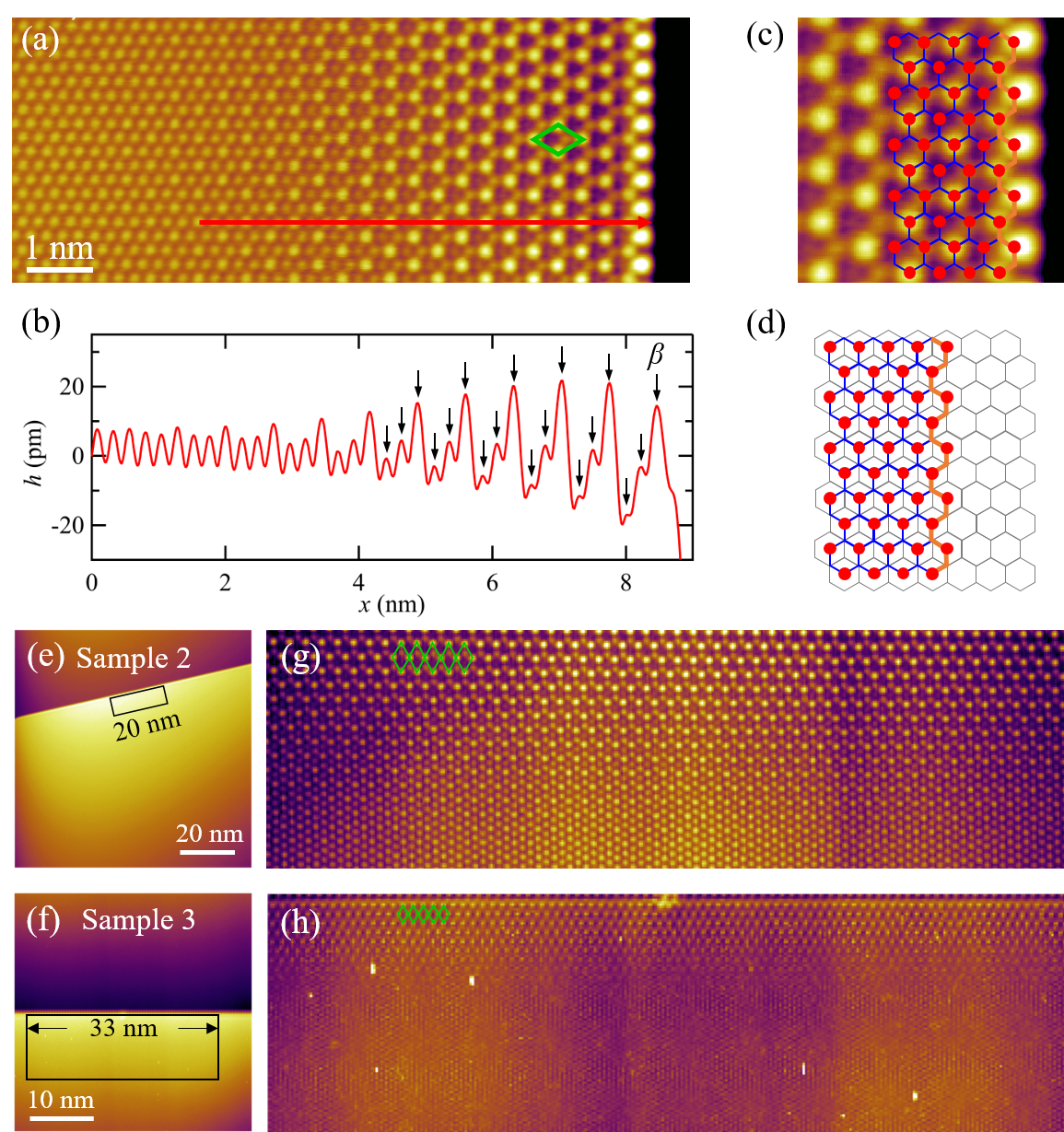}
\caption{(Color online) (a) Topographic image taken near a clean armchair edge. The \super superstructure is observed in this area, as marked by the green rhombuses. Tunneling condition: $V=+0.2$~V, $I=1$~nA. (b) Line profile taken along an atomic row of $\beta$ site marked by the red line in (a). The atomic corrugation of $\beta$-site carbon atoms is marked by black arrows. (c) Zoom-in topographic image on the upper terrace of the armchair edge, where the atomic sites are specified by the schematic. The red dots represent the $\beta$-site carbon atoms. (d) The atomic lattice of armchair edge. (e)(f) Homogeneous armchair edges observed on different samples. (g)(h) Zoom-in topographic images of upper terraces of the edges shown in (e)(f). The \super superstructure extends over 30~nm along the edge without interruption. The green rhombuses mark the unit cells of the superstructure.}
\label{Gr_TopoArmchair}
\end{figure*}

\begin{figure*}[htb]
\centering
\includegraphics[width=\textwidth]{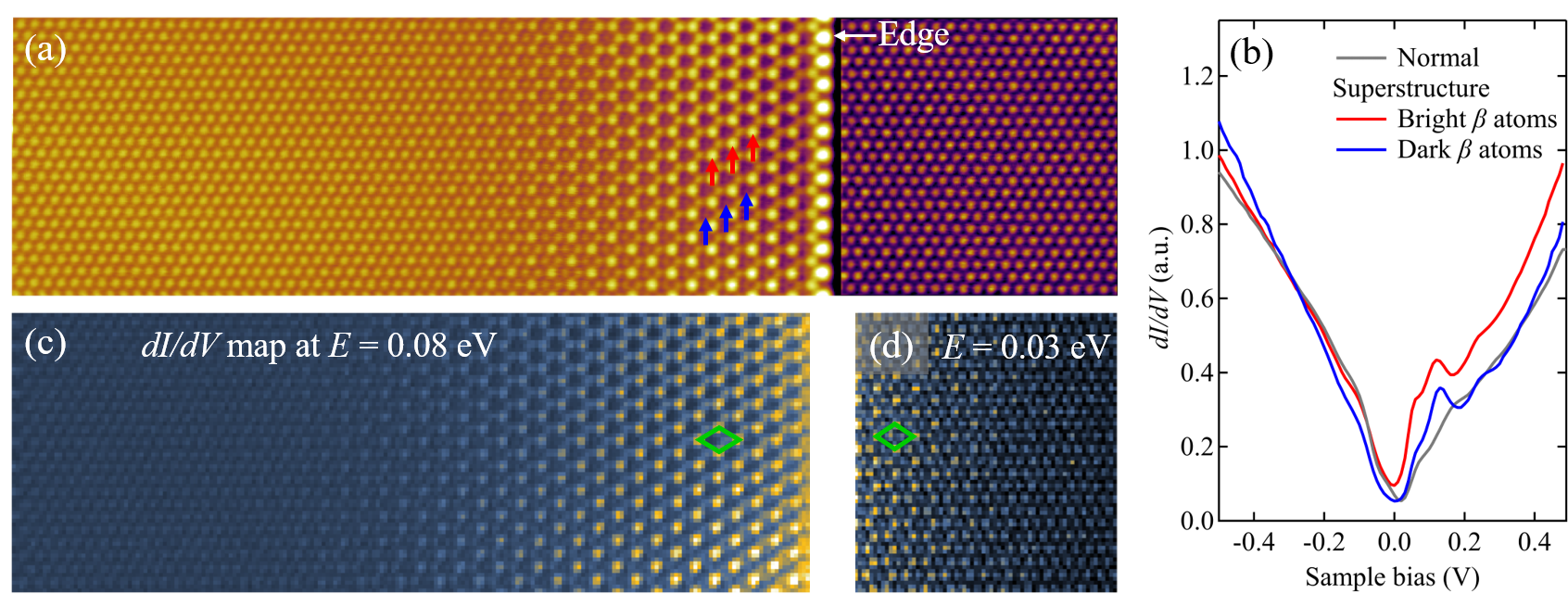}
\caption{(Color online) (a) Topographic image taken in the same area of Fig.~\ref{Gr_TopoArmchair}(b). No superstructure is observed in the lower terrace (right). (b) Individual $dI/dV$ spectra taken in different areas. (c-d) $dI/dV$ maps taken on the upper and lower areas, respectively.}
\label{Gr_dIdV}
\end{figure*}

In Fig.~\ref{Gr_TopoArmchair}, we present atomically resolved topographic images measured in the upper terraces near clean and homogeneous segments of armchair edge, which reveals that the \super superstructure is intrinsic to the armchair step edge of graphite. Fig.~\ref{Gr_TopoArmchair}(a) shows a typical region containing a clean part of armchair edge taken at $+0.2$~V. The color scale is adjusted to highlight the atomic corrugation on the upper terrace. Clearly, only the \super superstructure is visualized, as denoted by the green rhombuses. In Fig.~\ref{Gr_TopoArmchair}(b), the line profile taken along an atomic row of $\beta$ site [red line in Fig.~\ref{Gr_AtomicFlat}(a)] shows that the appear heights of three adjacent $\beta$-site atoms are all different, which further substantiates the existence of the \super superstructure. The superstructure extends over 6-7~nm from the edge. Figs.~\ref{Gr_TopoArmchair}(c)(d) show the detailed atomic lattice on the edge. Among the outermost carbon atoms, those at $\beta$ site appear as bright spots. In contrast, the $\alpha$-site atoms are not visible. The uniqueness of the $\beta$-site atoms on the edge probably results from the existence of dangling bonds at $\beta$ sites, as there are no carbon atoms in the adjacent layers that are aligned with them in $z$ direction.

The \super superstructure being intrinsic to the clean and homogeneous armchair edge is reproduced on multiple samples and robust to an elevated temperature as high as 48~K. Fig.~\ref{Gr_TopoArmchair}(e) shows a long clean armchair edge on the surface of a new piece of HOPG (denoted as Sample 2). The atomically resolved topographic image in Fig.~\ref{Gr_TopoArmchair}(g) indicates that the clear \super superstructure pattern extends over $\sim20$~nm along the edge. A third HOPG sample (Sample 3) was cleaved and measured at $\sim48$~K. Similarly, near the defectless armchair edge, the \super superstructure pattern persist longer than 30~nm along the edge, as shown in Figs.~\ref{Gr_TopoArmchair}(f)(h). In all these cases, no honeycomb superstructure is observed near clean and homogeneous armchair edges.

Our STS results obtained in the vicinity of armchair edge reveal that the superstructure observed in the topographic images originates from the unusual spatial distribution of LDOS. Fig.~\ref{Gr_dIdV}(a) shows a topographic image in the same area as Fig.~\ref{Gr_TopoArmchair}(a). The color scale of the lower terrace (the right side) of the edge is adjusted, so that the $\beta$-site carbon atoms in this area are also visible. As shown, they present a hexagonal lattice without clear superstructure patterns. Fig.~\ref{Gr_dIdV}(b) shows the typical individual $dI/dV$ spectra taken at selective locations. The red $dI/dV$ spectra was measured on top of pronounced $\beta$-site atoms near the edge [marked by red arrow in Fig.~\ref{Gr_dIdV}(a)]. It has higher LDOS at positive bias compared with the gray curve taken in the normal region far from the edge. This is possibly due to the positive interference of plane waves from the potential barrier created by the edge. The presence of the superstructure is characterized by the LDOS contrast between the pronounced atoms and the faint atoms at $\beta$ site [marked by the blue arrows in Fig.~\ref{Gr_dIdV}(a)]. The individual $dI/dV$ spectrum on pronounced atoms (red) and that on faint atoms (blue) do not overlap with each other within the energy range ($-0.5$~eV, $0.5$~eV), which suggests the superstructure would influence the electronic structure over a wide energy range. Indeed, our STS imaging measurement on the upper terrace of the edge shows a \super superstructure pattern from $-0.5$ to $0.5$~eV except at around $-0.27$~eV~\cite{supplimentary}. Across $-0.27$~eV, a phase shift of superstructure is observed, where the bright atoms become dark and vice versa. Consistently, the $dI/dV$ spectra taken on two types of $\beta$-site atoms in Fig.~\ref{Gr_dIdV}(b) intersect at that energy. Fig.~\ref{Gr_dIdV}(c) shows an exemplary $dI/dV$ map at $+0.08$~eV, where the superstructure pattern clearly appears. To our surprise, the superstructure pattern can also be observed in the $dI/dV$ map of the lower terrace, although it is much weaker than that on the upper terrace, as demonstrated in Fig.~\ref{Gr_dIdV}(d). It indicates that the potential barrier created by the step edge has a weaker influence on the electronic properties of electrons on the lower atomic layer. To our knowledge, the superstructure has never been observed on the lower terrace below the step edge in the prior reports. Another feature appearing in the $dI/dV$ spectra is a LDOS peak at $\sim0.12$~eV above the Fermi level ($E_F$) emerging in the superstructure region. This feature is reproducible near the different edges on multiple samples, and is consistent with prior STM studies~\cite{Niimi2006}.

\begin{figure}[t]
\centering
\includegraphics[width=\columnwidth]{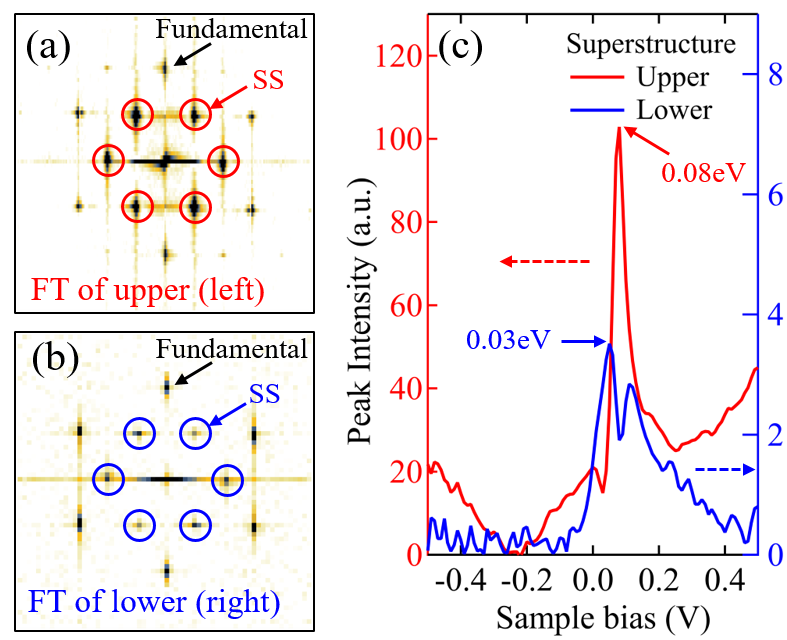}
\caption{(Color online) (a) FT of $dI/dV$ map in Fig.~\ref{Gr_dIdV}(c). The corresponding energy is $E=0.08$~eV, at which the superstructure (SS) is the most pronounced on the upper terrace. (b) FT of $dI/dV$ map in Fig.~\ref{Gr_dIdV}(d). The corresponding energy is $E=0.03$~eV, at which the superstructure is the most pronounced on the lower terrace. (c) The intensities of SS peaks in FT maps in (a-b) as functions of the energy.}
\label{Gr_FT}
\end{figure}

To quantitatively study the intensity of superstructure pattern in the STS results, we performed Fourier transform (FT) on the $dI/dV$ maps. Figs.~\ref{Gr_FT}(a)(b) show the FT of $dI/dV$ maps shown in Figs.~\ref{Gr_dIdV}(c)(d). The outer six peaks corresponds to the fundamental hexagonal lattice of $\beta$-site atoms. The inner six peaks marked by the red and blue arrows correspond to the \super superstructure patterns. Fig.~\ref{Gr_FT}(c) plots the intensities of superstructure peaks as functions of the energy. On the upper terrace of the edge, the peak intensity of superstructure is larger than zero in the whole energy range of ($-0.5$~eV, $+0.5$~eV) except a small energy interval around $-0.27$~eV, which is consistent with our direct observation in the $dI/dV$ map. On the other hand, the peak intensity of superstructure on the lower terrace takes a nonzero value only in a narrow energy range of 0 to $+0.2$~eV. Note that its highest value at $+0.03$~eV is $\sim25$ times smaller than that on the upper terrace at $0.08$~eV, which is consistent with the much weaker interlayer hopping energy compared to the intraplane hopping energy~\cite{CastroNeto2009}.

\begin{figure}[t]
\centering
\includegraphics[width=\columnwidth]{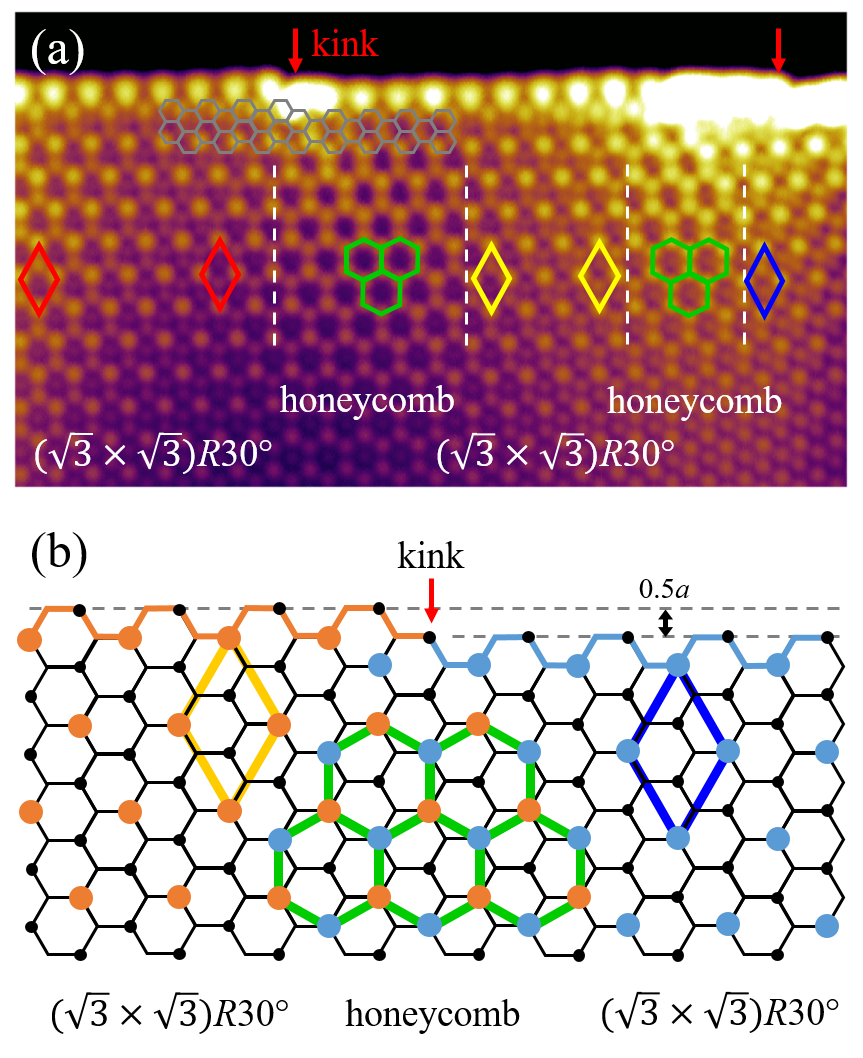}
\caption{(Color online) (a) Topographic image of armchair step edge with a mixture of the \super and honeycomb superstructures. The rhombuses and honeycombs illustrate the superstructures. The honeycomb superstructure is due to overlap of two set of intrinsic \super superstructure with a phase shift. (c) Lattice structure showing that the honeycomb superstructure emerges in the middle of armchair edges with a lattice offset. It is a result of overlapping of two sets of \super superstructure with a phase difference.}
\label{Gr_Mixture}
\end{figure}

So far, our experiments have revealed that only the \super superstructure emerges near the linear and defectless monoatomic armchair-type step edge, while no honeycomb superstructure is observed in the vicinity of such edge. However, we found that the honeycomb superstructure can exist if the armchair edge is not perfectly straight. In the region shown in Figs.~\ref{Gr_Mixture}(a), there are two kinks on the edge marked by the red arrows. As demonstrated by the gray mesh of honeycomb lattice, each kink is a very short zigzag-type edge located at the boundary of two fragments of armchair edge. The atomic row of terminating carbon atoms on the two sides of the kink has an offset of half the lattice constant perpendicular to the edge. The atomic lattice near the kinks is not perfectly linear, possibly because of the strain effect on the edge carbon atoms. Interestingly, the \super and honeycomb superstructures emerge alternatively, as demonstrated by the colorful rhombuses and honeycombs in Fig.~\ref{Gr_Mixture}(a) as a guide to the eye. Note that right under a homogeneous part of armchair edge is always the \super superstructure, while the honeycomb superstructure is visible near the kink between two parts of the armchair edge. Herein the coexistence of the two different types of superstructures can be understood phenomenologically as follows. In the case that a kink separates two segments of armchair edge with an offset of an atomic row, each part of armchair edge gives rise to an interference pattern (orange and blue dots) manifested as the \super superstructure (blue and orange rhombuses). With the equivalent apparent height at the antiphase boundary, the superposition of the two superstructure patterns creates a honeycomb superstructure (large green honeycomb), as illustrated in Fig.~\ref{Gr_Mixture}(b).

These kinks are not rare along the edge in a large scale. If the step edge of graphite or the edge of graphene is not clean or atomically smooth, it is very challenging to identify these structural imperfections with lack of clear atomic resolution in the vicinity of the edge. With many kinks, a visually straight armchair edge can still gives rise to a coexistence of two types of superstructures, as demonstrated in Fig.~\ref{Gr_Mixture}. In addition to the structural imperfection, chemical imperfection such as adsorbates may also contribute to the creation of more complicated superstructure. Future studies will address this issue.

In summary, our experiments reveal that the \super superstructure is the intrinsic superperiodic pattern of the clean and ideal armchair step edge of graphite. We visualized the \super superstructure not only on the upper terrace but also on the lower terrace, which indicates the electronic influence of the step edge as a potential barrier on both the first and second atomic layers. The honeycomb superstructure sandwiched by the \super superstructure is also observed near a structural imperfection that separates two parts of armchair edge with an offset. Such honeycomb superstructure can be viewed as superposition of two sets of intrinsic \super superstructures. This finding reconciles the different experimental observations about superstructure patterns in prior STM studies.

We thank Jixia Dai and Chen Chen for helpful discussions. This work was supported by NSF Grants No. DMR-1506618.

\section{Appendix: Identification of $\beta$-site carbon atoms via phase analysis}
\begin{figure}[t]
\centering
\includegraphics[width=\columnwidth]{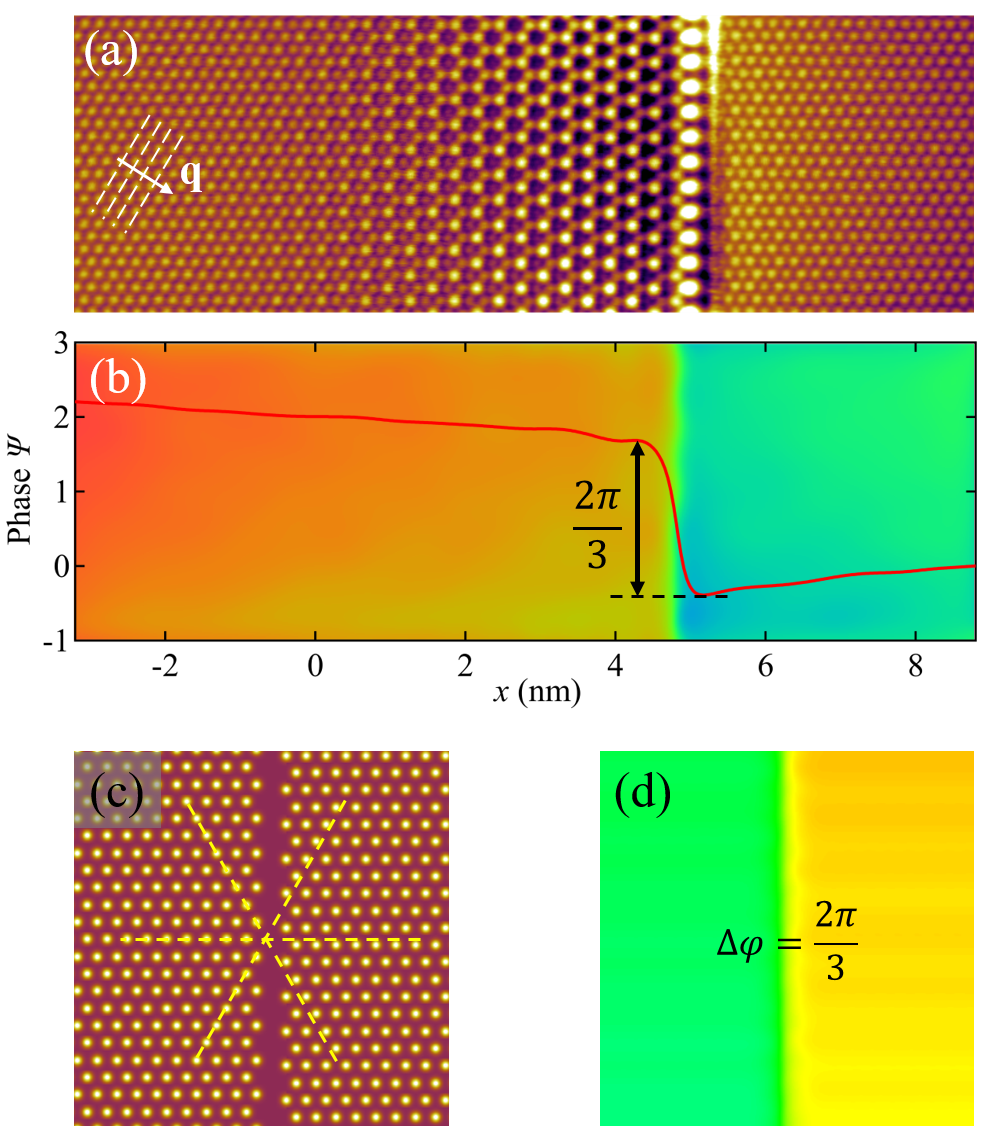}
\caption{(Color online) (a) Topographic image across the armchair edge, which is taken in the same area of Fig.~\ref{Gr_TopoArmchair}(a). $\mathbf{q}$ marks the direction in which the phase of lattice is calculated. (b) Phase map and line profile of phase taken perpendicular to the edge. The phase difference between the upper and lower terraces is $2\pi/3$. (c) Ideal hexagonal lattice with a phase shift of $2\pi/3$ at the antiphase boundary that simulates the $\beta$-site carbon atoms of two adjacent layers. (d) Corresponding phase map of (c).}
\label{Gr_phase}
\end{figure}

Via quantitative analysis, we confirm that the hexagonal lattice visualized in the topographic image taken at $+0.2$~eV is composed of $\beta$-site carbon atoms. The atomic row can be regarded as a plane wave in the direction of reciprocal lattice vector with a magnitude $q=2\pi/d$, where $d$ is the distance between two adjacent atomic rows, as demonstrated in Fig.~\ref{Gr_phase}(a). The direction and magnitude of $q$ can be accurately obtained in the FT of topographic image. The local phase of lattice associated with $q$ in the real space can then be extracted through FT analysis~\cite{Lawler2010}. The lattices of $\alpha$ site of two adjacent layers have the same phase, because they overlap in $z$ direction. In contrast, the lattices of $\beta$ site of two adjacent layers are not aligned with each other, and have a phase difference of $2\pi/3$. As shown by the phase map and line profile of phase in Fig.~\ref{Gr_phase}(b), the lattices on the two sides of the edge show a phase shift of $\Delta\phi = 2\pi/3$, which indicates the atoms visualized are at $\beta$ site. We also simulated the ideal $\beta$-site atomic lattices of two adjacent layers of graphite and calculated the phase map with the same analysis. As shown in Figs.~\ref{Gr_phase}(c)(d), the phase shift across the antiphase boundary equals $2\pi/3$, which agrees with our experimental results.

%
\end{document}